\documentclass[fleqn,usenatbib]{mnras}

\usepackage{mathptmx}

\usepackage[T1]{fontenc}
\usepackage{ae,aecompl}

\usepackage{graphicx,psfig}	
\usepackage{amsmath}	
\usepackage{amssymb}	

\newcommand{\mjyb}{mJy beam$^{-1}$\,}
\newcommand{\ugmrt}{uGMRT}

\title[Spectral curvature in A4038]{A study of spectral curvature in the radio relic in Abell 4038 using the 
\MakeLowercase{u}GMRT}

\author[Kale et al.]{
Ruta Kale,$^{1}$\thanks{E-mail: ruta@ncra.tifr.res.in}
Viral Parekh,$^{2}$
and K. S. Dwarakanath$^{2}$
\\
$^{1}$National Centre for Radio Astrophysics, Tata Institute of Fundamental 
Research, Post Bag 3, S. P. Pune University Campus, \\Ganeshkhind,
Pune 411007, Maharashtra, India\\
$^{2}$Raman Research Institute, C. V. Raman Avenue, Sadashivanagar, 
Bangalore 560 0 80, Karnataka, India\\
}

\date{Accepted XXX. Received YYY; in original form ZZZ}

\pubyear{2015}

\begin{document}
\label{firstpage}
\pagerange{\pageref{firstpage}--\pageref{lastpage}}
\maketitle

\begin{abstract}
The remnant radio galaxies in galaxy clusters are important sources 
of seed relativistic electron population in the intra-cluster medium (ICM). 
Their occurrence and spectral properties are poorly studied. In this work we 
present a broadband study of the radio relic in the galaxy cluster Abell 4038 
using the Upgraded Giant Metrewave Radio Telescope (uGMRT). We present the 
uGMRT images in the bands 300 - 500 MHz and 1050 - 1450 MHz 
having rms noise $70\,\mu$Jy beam$^{-1}$ and $30\,\mu$Jy beam$^{-1}$, 
respectively, that are the deepest images of this field so far. A spectral 
analysis of the relic over 300 - 1450 MHz using images in sub-bands scaled to 
have constant fractional bandwidths to achieve a closely matched uv-coverage was 
carried out.
The 100 kpc extent of the relic is divided into Loop, Arc, Bridge and  
North-end. The Loop has a steep spectral index 
of $\alpha=2.3\pm0.2$ ($S_{\nu}\propto\nu^{-\alpha}$). The 
North-end has ultra-steep spectra in the range $2.4 - 3.7$. The Arc is found to 
skirt a curved region seen in the \emph{Chandra} X-ray surface brightness image 
and  the highest spectral curvature in it reaches $1.6\pm0.3$. We interpret the 
morphology and spectral properties of the relic in the scenario of an 
adiabatically compressed cocoon from the past activity of the Brightest Cluster 
Galaxy in the cluster. A comparison of the properties of the A4038 relic with a 
sample of 10 such relics is discussed. 
\end{abstract}




\section{Introduction}
\label{intro}
Clusters of galaxies contain among the largest pools of baryons in the 
Universe called the intra-cluster medium (ICM). The ICM contains 
thermal gas at temperatures $10^7 - 10^8$ K and is also known to be magnetized 
at 
$\mu$G levels. About a third of the most massive clusters, radio synchrotron 
sources of extents of 
Mpc, co-spatial with the X-ray emission from the thermal 
gas or at the periphery are found \citep[e. g. ][a review]{bru14}. 
These provide evidence for the 
presence of electrons with relativistic energies in the ICM. Since diffusion 
time of such electrons from a single source across the cluster is far 
longer than their radiative lifetimes, processes of re-acceleration are invoked 
\citep{jaf77}.
The cluster-wide diffuse synchrotron sources called the radio halos are 
believed to originate from in-situ re-acceleration of electrons via turbulence 
injected during cluster mergers 
\citep[][]{1987A&A...182...21S,pet01,bru04,bru11,bru16}. The arc-like radio 
sources at cluster peripheries are proposed to originate in the re-acceleration 
of mildly relativistic electrons at cluster merger shocks \citep[e. 
g.][]{ens98}. The turbulent re-acceleration and diffusive shock acceleration at 
the weak shocks in the ICM are both inefficient processes requiring, 
 a seed population of mildly relativistic electrons \citep[e. 
g.][]{Kang2007,Bruggen2012,pin13,pin17}.

The sources of seed relativistic electrons could be the secondary electrons 
produced in the hadronic collisions \citep{bru11} or those injected by the  
cluster galaxies \citep{bru14}. The Active Galactic Nuclei (AGN) and radio 
galaxies inject relativistic plasma via jets during the active phase of 
the central supermassive blackhole. However the jets and lobes of radio 
galaxies will fade 
on the timescales of 
$10^7-10^8$ years unless there are mechanisms at work that re-energize them 
\citep[e. g.][]{1987A&A...182...21S}. 
Adiabatic compression of remnant radio cocoons due to shocks and disturbances in 
the ICM has been proposed to revive the radio emission \citep{ens01}. 
 Simulations of this process predict that the compressed cooon will be shred 
into filaments allowing dispersal of the plasma 
\citep{2002MNRAS.331.1011E}. A mechanism of gentle re-acceleration also has 
been proposed recently 
\citep{2017NatAs...1E...5V}. Such remnants that show revival from their fading 
phase have also been referred to as radio ``phoenixes'' \citep{kemp04} in the 
literature.

In this work we focus on remnant relics that could be important sources 
of seed relativistic electrons in the ICM. The detection of such remnants in 
clusters is challenging due to their low brightness and rare 
occurrence due to short radiative lifetimes. A sample of 13 
such relics has been presented in \citet{fer12} where these are classified as 
``roundish'' relics to distinguish them from the arc-like and elongated relics 
associated with merger shocks. A further larger sample of ``phoenix'' 
candidates has been presented in \citet{nuza17} and recently a faint 
sub-sample of these relics has been presented by \citet{Wilber2018}. Integrated 
spectra of such relics over broad ranges of frequencies have been measured for 
some of the relics \citep{sle01,wee11b,cohen11,kaldwa12}. Spectral index 
studies of a few relics with good resolution across the relic have also been 
carried out by combining multi-frequency radio data from one or more radio 
telescopes and have revealed the complex morphologies and spectra 
 and allowed to estimate the life cycle of the 
relativistic plasma in them 
\citep[e. g.][]{cohen11,kaldwa12,2015A&A...583A..89S,2017A&A...600A..65S}.

Here we present a broadband study of a remnant radio relic towards the 
galaxy cluster Abell 4038 (A4038, hereafter) using the Upgraded Giant Metrewave 
Radio Telescope (uGMRT).
The uGMRT observations resulting in the deepest images towards this source are 
presented and used to measure spectral curvature across the extent of the 
relic. The radio morphology, spectra and the X-ray surface brightness map of 
the cluster together point to a scenario in which the relic originated from the 
compression of a radio cocoon left by the central galaxy. A sample of such 
relics is presented and the A4038 relic in comparison with others is discussed 
and the importance of spatially resolved wideband observations of these sources 
is emphasized. The paper 
is organized as follows: Sec.~\ref{a4038} describes the cluster Abell 4038 and 
the relic in it. The 
observations and data analysis are described in Sec.~\ref{obs}. The uGMRT 
images are presented in Sec.~\ref{ugmrtim}. The spectral curvature analysis is 
described in Sec.~\ref{curv}. The results are discussed in 
Sec.~\ref{discussion}. The summary and conclusions are
presented in Sec.~\ref{conclusions}.

We have used $\Lambda$CDM cosmology with $H_0$=$70$ km s$^{-1}$ Mpc$^{-1}$, 
$\Omega_\Lambda$=$0.73$ and $\Omega_m$ = $0.27$ in this work.

\section{Abell 4038}\label{a4038}
Abell 4038 (also known as Klemola 44) is a galaxy cluster of richness class 2 
at a redshift of 0.03 having a bolometric X-ray luminosity of 
$(1.900\pm0.025) \times 10^{44}$ erg s$^{-1}$ \citep{mittal11}. The 
properties of the cluster are listed in Table~\ref{clusprop}.
\begin{table}
\centering
\caption{\label{clusprop}Properties of Abell 4038.}
\begin{tabular}{ll}
\hline\noalign{\smallskip}
RA$_{\rm J2000}$ & 23h47m43.2s  \\
\smallskip
DEC$_{\rm J2000}$ &-28$^{\circ}$08$^\prime$29$^{\prime\prime}$ \\
\smallskip
Redshift$^\dag$ & $0.02819\pm0.00055$\\
\smallskip
$kT^\dag$ & $2.69\pm0.43$ keV\\
\smallskip
$L_{[0.01-40]keV}$ $^{++}$ & $(1.900\pm0.025) \times 10^{44}$ erg s$^{-1}$\\
\smallskip
M$^\ddag$ & $1.5\pm0.1 \times 10^{14}$ M$_{\odot}$\\
\noalign{\smallskip}
\hline\noalign{\smallskip}
\end{tabular}\\
$\dag$ \citet{2011MNRAS.410.1797S}
$++$ \citet{mittal11}
$\ddag$ \citet{2016A&A...594A..27P}
\end{table}

A diffuse steep spectrum source towards this cluster was discovered by 
\citet{1983PASAu...5..114S}. In a further study at 1.4 GHz they reported a 
radio relic of size 56 kpc, with polarization at 1.4 GHz of $4.6\pm2.3\%$ 
and a spectral index of $3.1$\footnote{The spectral index $\alpha$ of the 
synchrotron spectrum is defined as, $S_\nu\propto\nu^{-\alpha}$, where $S_\nu$ 
is the flux density of the source at the frequency $\nu$.}
\citep{sle01}. Further high resolution multi-frequency study of the relic was 
presented by \citet{kaldwa12}. 
The Giant Metrewave Radio Telescope (GMRT) observations at 
frequencies 150, 235 and 610 MHz revealed new steep spectrum regions of the 
radio relic of extent 100 kpc and possibly up to 200 kpc.
The integrated spectrum of the relic was 
modeled in the framework of the theoretical model 
by \citet{ens01}. The relic was found to be consistent 
with the model of a radio galaxy lobe that is energized by adiabatic 
compression due to a passing shock wave. 
It was clear from this study 
that the relic showed complex morphological and spectral features that were 
not resolved. With the new uGMRT observations we have performed a spatially 
resolved spectral study of the radio relic.

\section{Observations and Data Reduction}\label{obs}
An upgradation of the receivers of the GMRT is presently being carried out. The 
 upgraded GMRT (\ugmrt) will have new receivers providing a 
near-seamless frequency coverage between $50 - 1500$ MHz 
\citep{Gupta2017}. 
The observations towards A4038 were carried out in the shared risk category 
during the Cycle 31. 
The data were recorded in 2048 frequency 
channels and two polarizations (RR, LL) with a sampling time of 8 seconds and 
observing duration of 8 hours in 
 the bands 300 - 500 and 1050 - 1450 MHz (Table ~\ref{obsdetail}). 
The flux calibrators, 3C48 and 3C147, were used to calibrate the bandpass and 
absolute flux density scale. The secondary calibrator, 0025-260, was used to 
calibrate the phases towards the target. 
\begin{table*}
\begin{center}
\caption{\label{obsdetail} Summary of the uGMRT observations.}
\begin{tabular}{cccccccc}
\hline\noalign{\smallskip}
Project Code&Date & Freq. Band& BW   & Freq. res.&Time &Beam&$\sigma_{rms}$ \\
         & &MHz  & MHz &kHz channel$^{-1}$& hr & $''\times''$, 
$^\circ$&mJy 
beam$^{-1}$\\
\hline\noalign{\smallskip}
31\_067     & 06 Feb. 2017 & 300 - 500 & 200 & 97.7 &8&$10.0\times5.0$, 
$14.1$  &0.07\\
31\_067     & 12 Mar. 2017 & 1050 - 1450 & 400 &195.3 & 8& $3.6\times1.74$, 
$37.6$  & 0.03\\
\hline\noalign{\smallskip}
\end{tabular}
\end{center}
\end{table*}

Data were analyzed using the AOFlagger 
\citep{2012A&A...539A..95O} 
and Common Astronomy Software Applications (CASA) packages 
\citep{2007ASPC..376..127M}. 
After removing non-working antennas, we ran AOFlagger with default 
parameters on the whole data set. It removed channels and time periods 
affected by radio freqeuncy interference (RFI). In both the frequency 
bands, it flagged $\sim$ 25\% data. For bandpass and complex gain calibration, 
we used the standard steps in CASA. The flux density of the flux calibrators 
was set according to \citet{2013ApJS..204...19P}. The AOFlagger was run on 
the calibrated visibilities. The target source 
(A4038) visibilities were then split after averaging 
of 10 channels. AOFlagger was used on these data to remove any 
remaining bad data before proceeding for imaging. The images were made using 
the final visibilities and standard steps of self-calibration. The 
multi-term multi-frequency synthesis mode (MT-MFS) in the CASA 
task clean was used to produce the final images. For the spectral index 
analysis we used a new method that ensured closely matched uv-coverage across 
the 0.402 - 1.4 GHz (Sec.~\ref{curv}).

\section{uGMRT images of A4038}\label{ugmrtim}
\begin{figure*}
 \begin{center}
\includegraphics[trim=0.cm 0.cm 0.cm 0.0cm,clip,width=8.5cm]{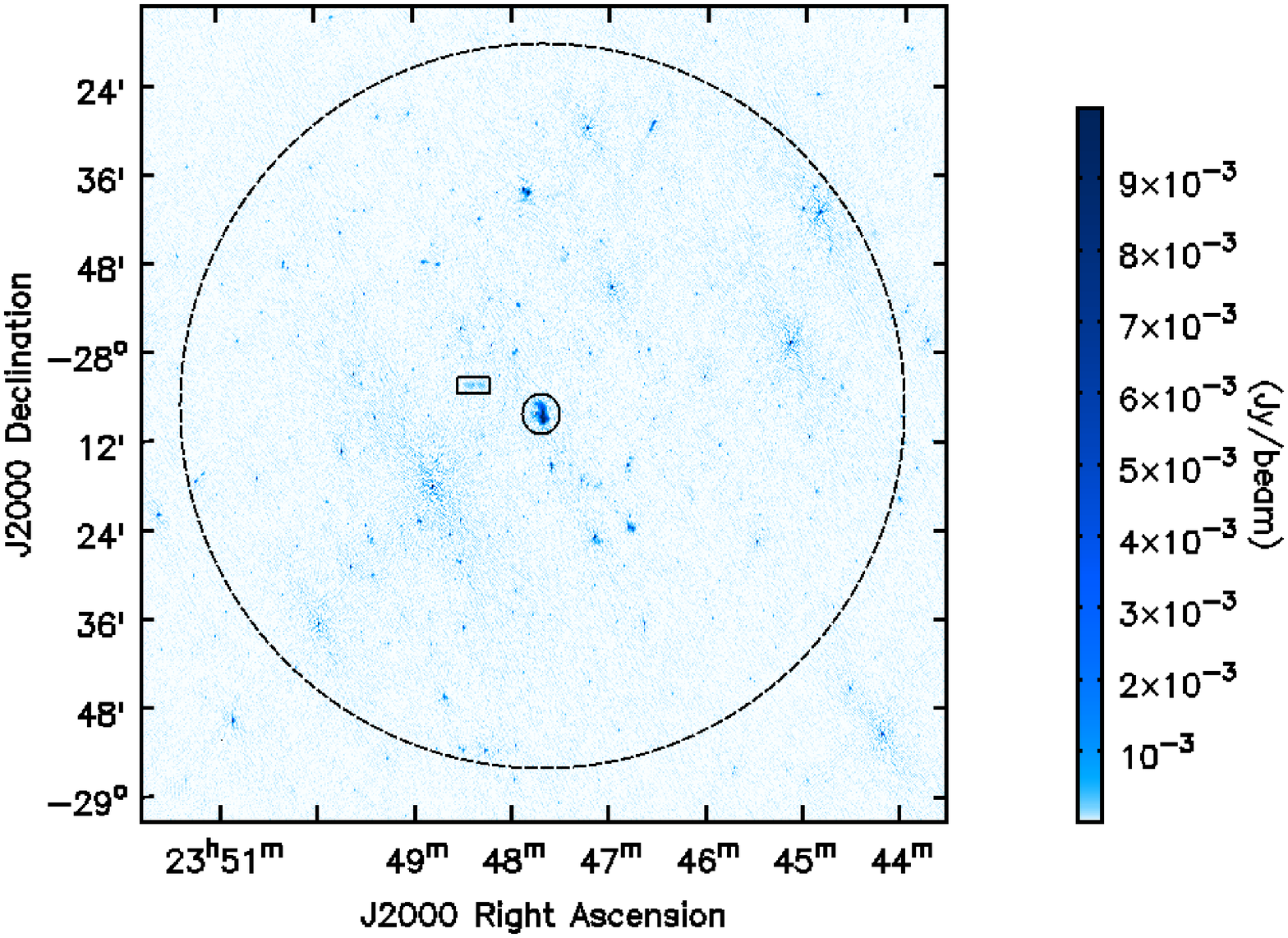}
 \hspace{1mm}
  \includegraphics[trim=0.cm 
 0.cm 0.cm 0.0cm,clip,width=8.5cm]{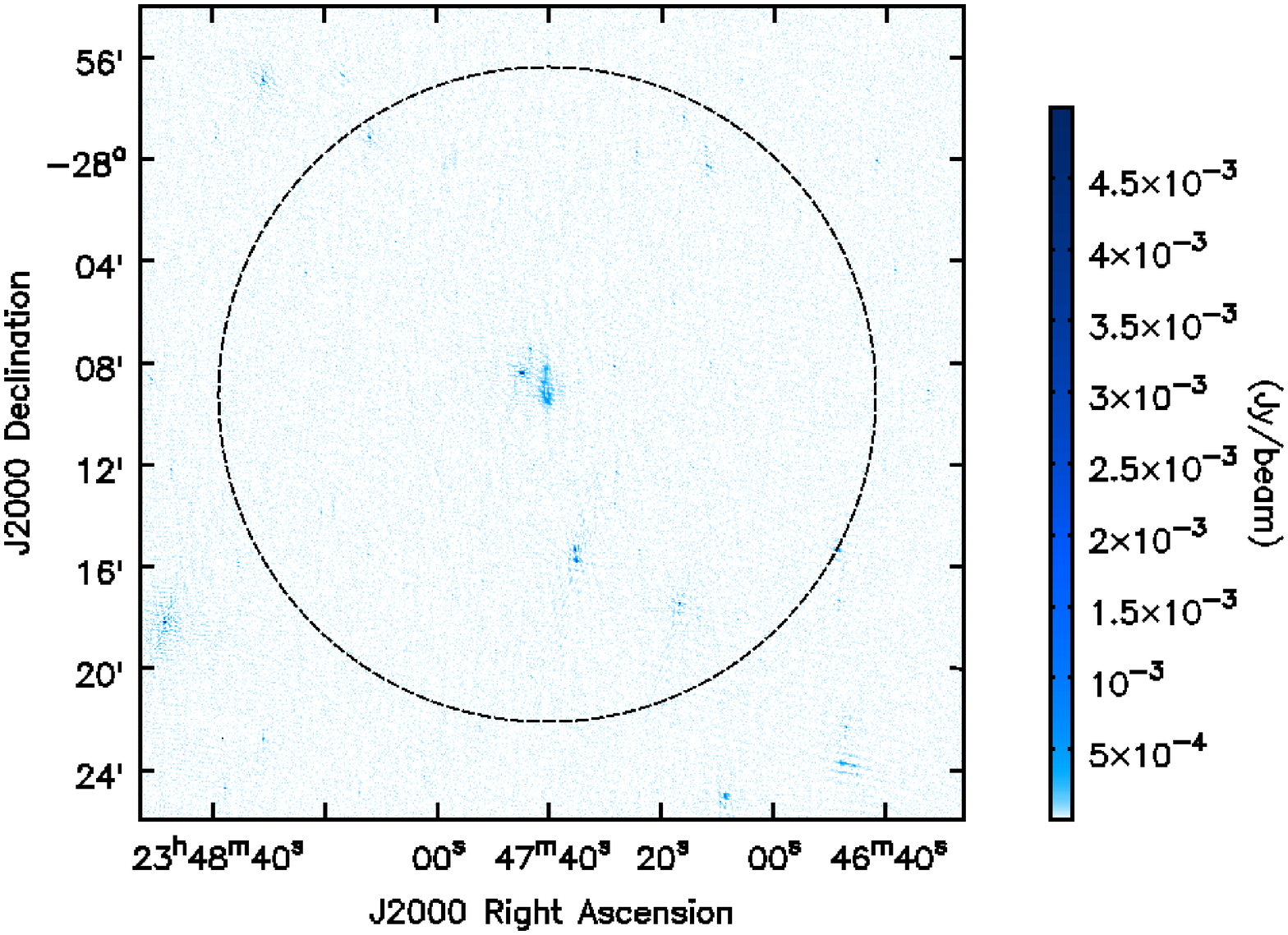}
\caption{Left:- The uGMRT 402 MHz image using data between 300 - 500 MHz. The 
rms is 0.07 \mjyb near the centre of the field. 
The central ellipse shows the relic in A4038 and the rectangle  
marks the detection of diffuse lobes of a radio galaxy. Right:- The uGMRT 1.26 
GHz image using data between 1050 - 1450 MHz. The rms 
is 0.03 \mjyb near the centre. 
The synthesized beams are reported in 
Table~\ref{obsdetail}. In both the panels, circles with diameters equal to the 
respective primary beam half power widths are shown.
\label{imgpl}}
\end{center}
\end{figure*}

The uGMRT images at the effective frequencies of 402 MHz and at 1.26 GHz are 
shown in Fig.~\ref{imgpl}. 
The radio relic in A4038 is at the centre of the field. The detection of 
diffuse steep spectrum lobes of a radio galaxy in the 402 MHz image is also  
marked (Appendix~\ref{app}). 

\begin{figure*}
 \begin{center}
\includegraphics[width=13cm]{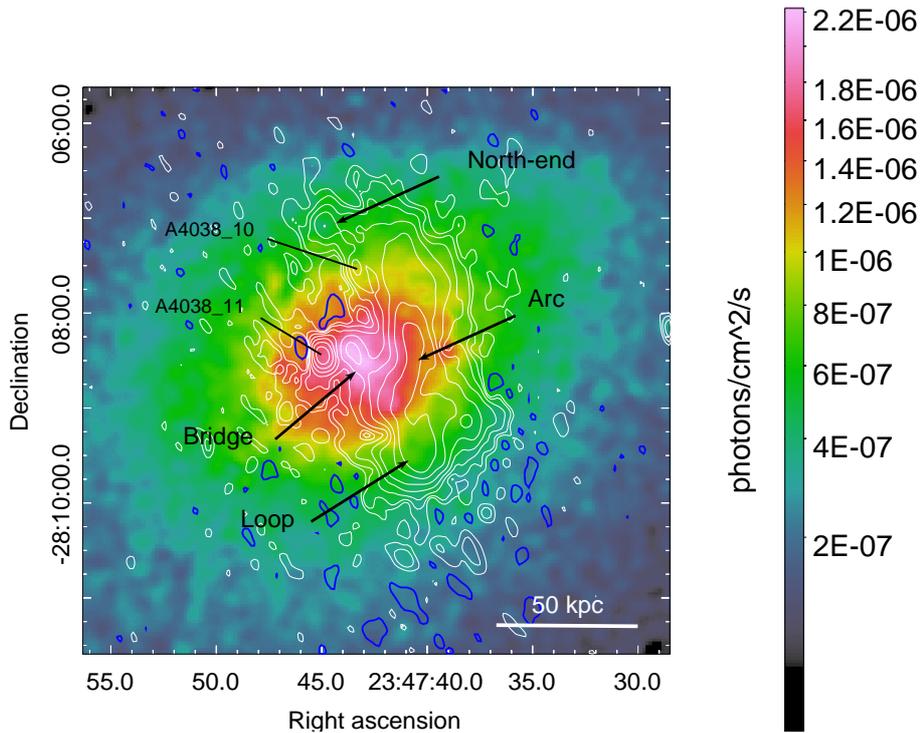}
\caption{The uGMRT 402 MHz image is shown in contours overlaid on the 
\emph{Chandra} $0.5 - 7$ keV X-ray image (OBSID 04992) in colour. The contour 
levels start at 
$\sigma\times[-3, 3, 6, 12, ...]$, where $\sigma= 0.07$ \mjyb is the rms noise. 
Positive contours are white and the negative are blue. The discrete sources are 
A$4038\_11$ and A$4038\_10$. The parts of the radio relic are labelled ``Arc'', 
``Loop'' and ``Bridge''.  \label{pxray}}
\end{center}
\end{figure*}

\begin{figure*}
 \begin{center}
\includegraphics[height=9cm]{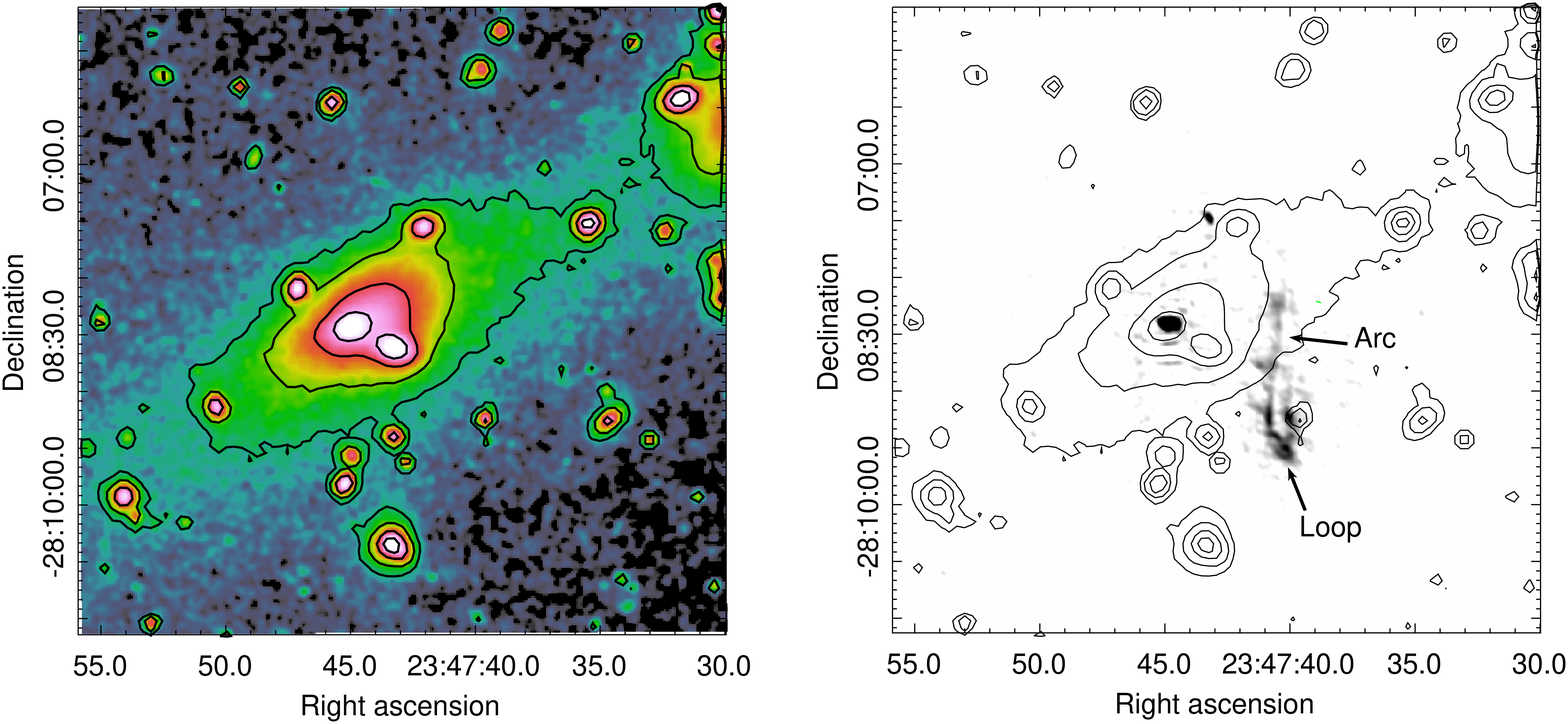}
\caption{Left:- Digitized Sky Survey POSS II R-band optical image {\bf of A4038} is shown in 
colours and contours. Right:- The 1.26 GHz image with resolution 
$3.6''\times1.47''$, p. a. $37.6^{\circ}$ is shown in grey scale and the 
optical 
contours same as those in the left panel are overlaid. The ``Loop'' and 
the ``Arc'' are labelled.
\label{lopt}}
\end{center}
\end{figure*}

The 402 MHz image of A4038, overlaid on the \emph{Chandra} X-ray image (OBSID 
04992, 0.5 - 7 keV) is 
shown in Fig.~\ref{pxray}. The different parts of the extended radio relic are 
labelled as the Loop, Arc, North-end and Bridge. The Loop and the Arc are 
well resolved at 1.26 GHz and fall to the west of the BCG and 
do not show any obvious connection to the galaxies in that region  
(Fig.~\ref{lopt}).
There are two discrete sources in the cluster region 
labeled A4038$\_11$ and A4038$\_10$ (Fig.~\ref{pxray}). The source A4038$\_11$ is 
associated with the brightest cluster galaxy (BCG) in the cluster as seen in 
the optical band (Fig.~\ref{lopt}). The second 
discrete source A4038$\_10$ is likely a distant source as no optical 
counterpart can be seen.

The overall morphology of the relic is 
filamentary. The Arc and the Loop are the brightest regions of the 
relic that are detected across all the wavelengths. 
The Loop is resolved at 1.26 GHz (Fig.~\ref{lopt}, right) and reveals the 
knot-like structure that makes the southern brightest region of the relic.
At lower frequencies a region labeled as the ``Bridge'' between  A4038$\_11$ 
and the Arc is detected (Fig.~\ref{pxray}). The Bridge is co-spatial with the 
brightest region in X-rays that is also elongated in the same direction as the 
Bridge. In addition, an extended part of the Arc 
towards the north  turns west and terminates at a brighter but, diffuse, region 
labeled the ``North-end''. An extension, further towards the north-west, that 
was speculated based on 240 MHz image by \citet{kaldwa12}, is not detected, and 
was an artifact. At 402 MHz the radio relic has an angular 
extent of $3'$ ($102$ kpc) in the north-south direction and a largest width of 
$1.7'$ ($58$ kpc) in the east-west. 

The flux densities of the discrete sources A4038$\_11$ and A4038$\_10$ were 
measured from the images made with the highest resolution in order to minimize 
the contamination due to the radio relic. A4038$\_11$ has a flux density of 
$51\pm2$ mJy at 402 MHz and $31\pm3$ mJy at 1.26 GHz and A4038$\_10$ has a flux 
density of $10\pm1$ mJy and $3.3\pm0.3$ mJy at the respective frequencies. 
The flux densities for these discrete sources are reported based on the 
Gaussian fit obtained to the detected source assuming a single component. 

\begin{figure*}
 \begin{center}
\includegraphics[trim=2.cm 
5cm 2.cm 4cm,clip,height=15cm]{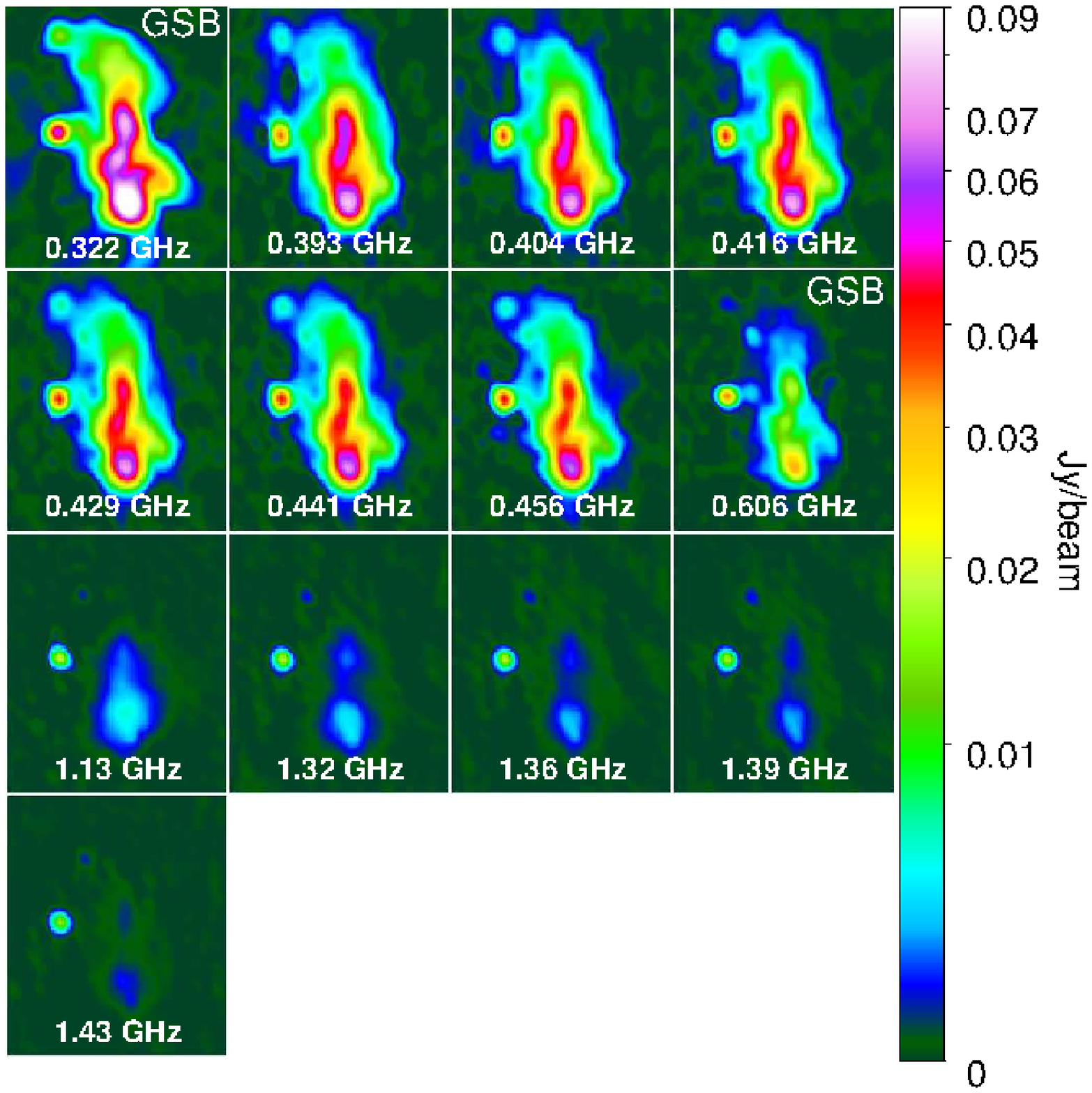}
\caption{The images of A4038 relic at each of the narrow 
bands used for the spectral analysis. The two with the 
legacy GMRT system are marked as ``GSB''. The remaining are 
from the data from the uGMRT. {The colour scale in the range  
0.00 - 0.09 Jy beam$^{-1}$ is common 
for all the panels.} All the images have been convolved to the 
common resolution of $10.4''\times10.4''$.\label{imgcollage}}
\end{center}
\end{figure*}

\section{Spectral curvature}\label{curv}
\subsection{Matched uv-data}
The wideband observations were used to analyze the continuous variation 
of the spectral index across the frequencies and across the spatial 
extent of the relic. The RFI typically severely affects the short baselines 
which are crucial for imaging extended sources. Since 
our focus of study is the extended radio relic, we chose the regions of 
bandwidth that had suffered minimal flagging at short baselines. This resulted 
in the choice of frequency ranges from 380 - 460 MHz in Band 3 and 1300 - 
1440 MHz in Band 5 for this analysis. Furthermore, to study the continuous 
spectral variation, it is important that our sampling in the uv-plane be  
similar in the frequency ranges that we consider.
We fixed a bandwidth of 11 MHz at 380 MHz 
 for a secure detection of the relic, and split the data into measurement sets 
with narrow bandwidths keeping the $\Delta \nu/\nu = 0.028$ constant over the 
range of frequencies between 380 - 1440 MHz. 
This ensures that the width of the uv-track is the same across the bands 
 to match the uv-coverage. This results in sub-band measurement sets with 
 increasing bandwidth as a function of frequency. In our case the 
bandwidths range from 11 MHz at the lowest to 40 MHz at the highest frequency 
end. These bandwidths thus represent the limits below which the spectrum is 
assumed to be straight. The images were separately made from each of the 
measurement sets using ``uniform'' weights (robust$ = -2.0$) for the 
visibilities. The uv-range while imaging is 
also restricted to the range, 200 - 31000 $\lambda$, which is the overlapping 
uv-range across the selected frequency bands. These images produced with 
closely matched uv-coverage were inspected and those with a poor rms noise were 
excluded from the analysis. 
 The final images were convolved to a common resolution of $10.4''\times10.4''$ 
and used for the spectral analysis. The images from individual sub-bands and 
those from the legacy GMRT are shown in Fig.~\ref{imgcollage}.
\subsection{Spectra}
\begin{figure}
\includegraphics[trim=2.cm 
1.cm 2.cm 1.cm,clip,width=8.5cm]{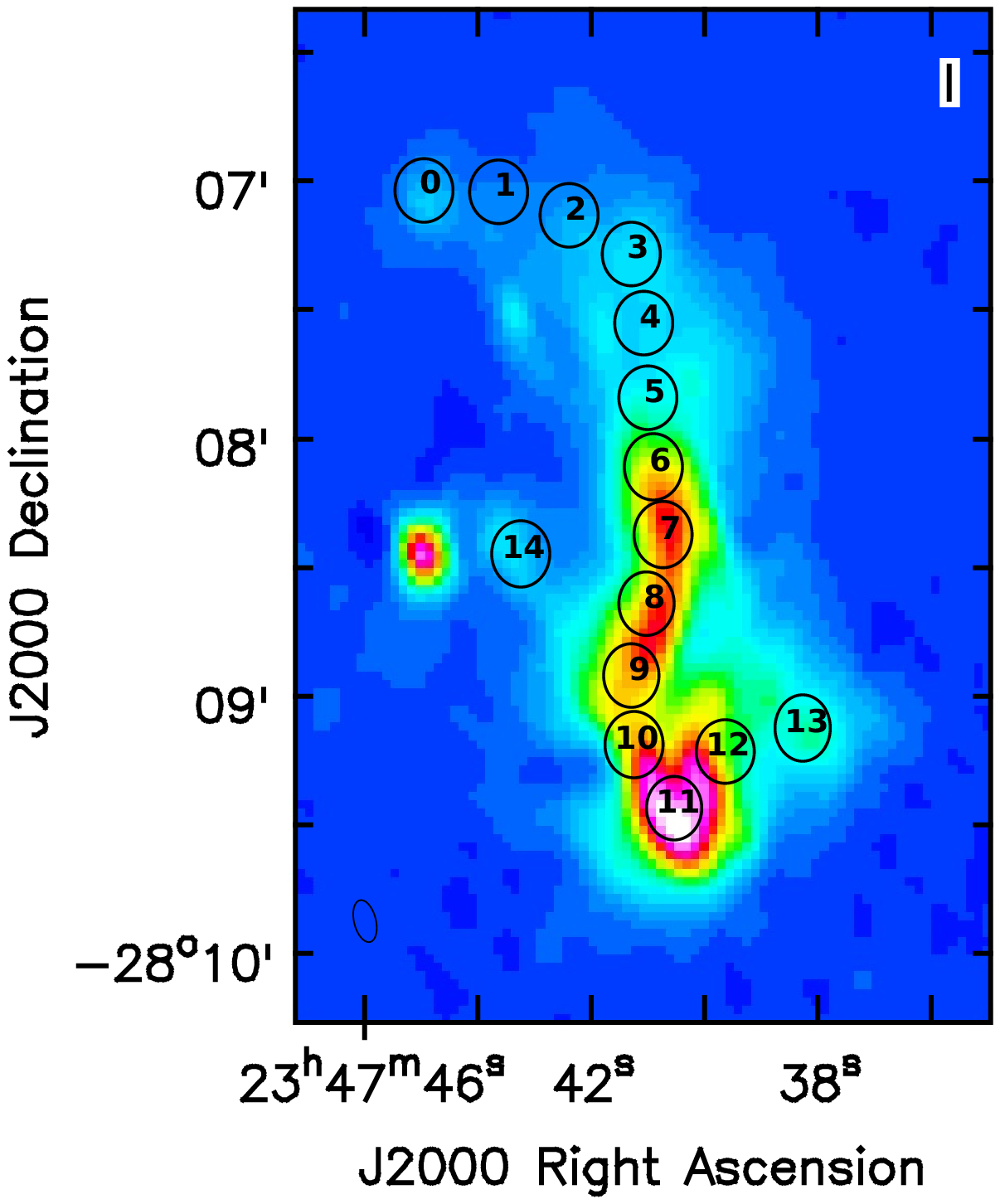}
\caption{The regions in which the spectral curvature analysis of the 
relic is carried out are marked on the 402 MHz image. \label{regions}}
\end{figure}

\begin{figure*}
\centering
\includegraphics[trim=1.0cm 
2.0cm 0.5cm 
2.0cm,clip,height=8.2cm]{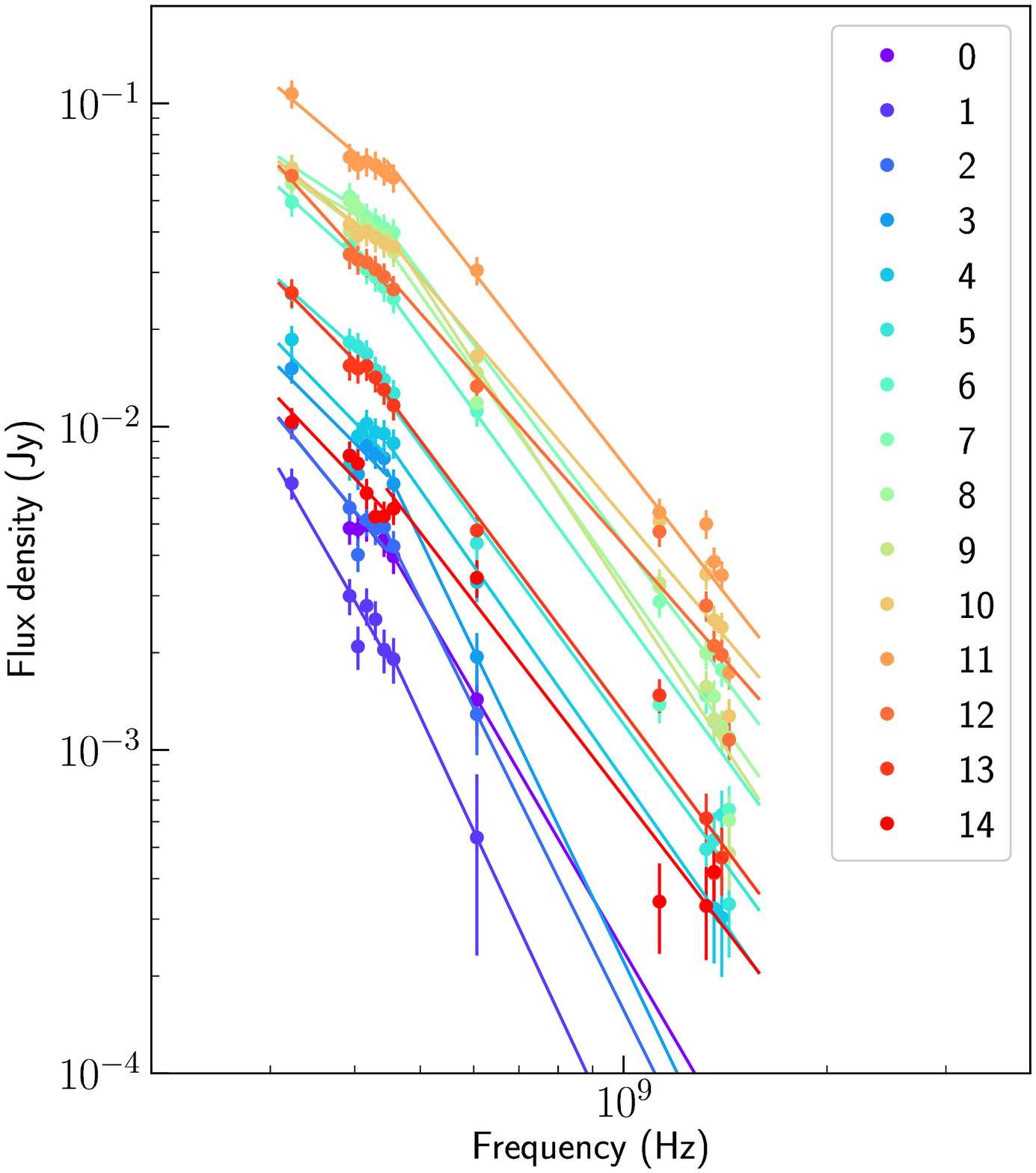}
\includegraphics[trim=0.5cm 
0.0cm 1cm 1cm,clip, height=8.0cm]{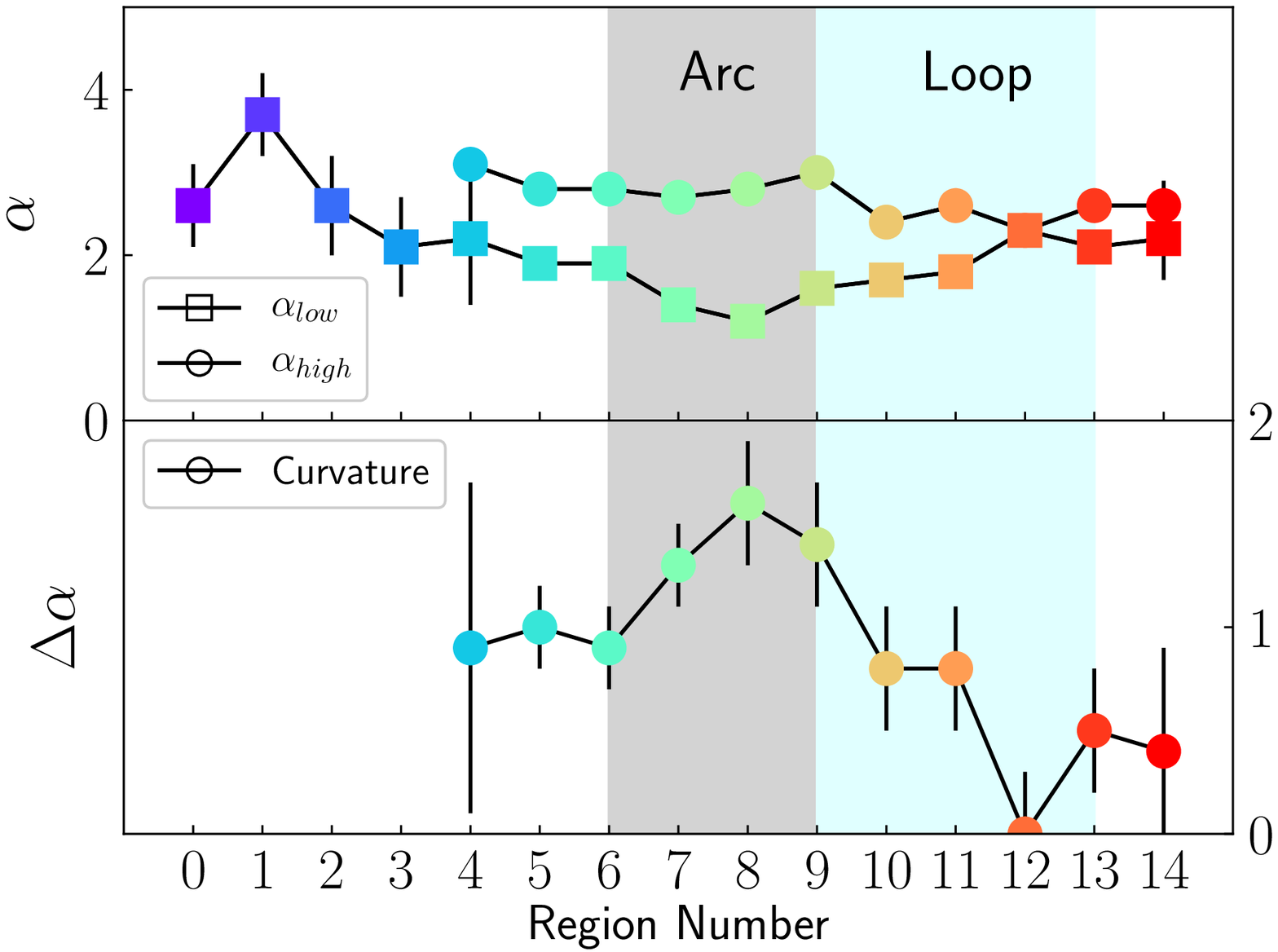}
\caption{Left:- The spectra of the regions marked in Fig.~\ref{regions} are 
plotted with a range of colors as shown in the legend. The best fit power-law 
spectra in the low and high frequency range are shown with solid lines. 
Right:- In the top panel, spectral indices $\alpha^{0.45}_{0.32}$ and 
$\alpha^{1.43}_{0.45}$ are plotted as a function of the region number shown on 
the x-axis. The difference between the two spectral indices (curvature) 
 as a function of the region number is shown in the bottom panel. 
The grey shaded region marks the region numbers corresponding to the Arc and 
the cyan shaded region marks the Loop. To the right of the Loop is the region 
14 corresponding to the Bridge. To the left of the Arc are regions in the 
northern part that are detected only at low frequencies and thus have no 
higher frequency spectral index and curvature measurements. The color scheme 
for the region numbers is the same as in the left panel. 
\label{spec}}
\end{figure*}

The relic was divided into regions of size $15''\times15''$ for spectral 
curvature analysis (Fig.~\ref{regions}). This size was chosen to be larger than 
the beam so that independent regions are used for spectral analysis. The 
selected regions cover the North end (0), the northern part of the relic (1 - 
5), the Arc (6 - 9) and the Loop (10 - 13). The region number 14 covers the 
Bridge. The flux densities in these regions were extracted from the images 
presented in Fig.~\ref{imgcollage}. For the regions 0 - 3 a single power-law 
was fit to the measurements at low frequencies. In the sub-band images at high frequencies,  
these regions were not detected due to their steep spectra. From region 4 to 14 
the a separate power-law fits 
were made in the frequency ranges 0.32 - 0.45 GHz and 0.45 - 1.43 GHz. The 
linear fits were of the form, 
\begin{equation}
 \rm{log \, S} = -\alpha_{\rm low} \rm{log\,} \nu + {\rm B}.
\end{equation}
For the high frequency range the derived parameters were $\alpha_{high}$ and 
$C$. These are listed in Table~\ref{fit}. The spectra for the regions are 
plotted in Fig.~\ref{spec} (left). 

The difference, $\Delta \alpha = 
\alpha_{\rm high} - \alpha_{\rm low}$ provides a measure of the spectral 
curvature and is plotted for each of the regions in Fig.~\ref{spec}. The 
curvature is positive in all the regions with variation between 0.0 - 1.6.
Along the Arc from north to south, the curvature increases from 1 to 1.6 and 
then decreases in the Loop. The regions 0 - 3 need high sensitivity 
measurements above 0.6 GHz to quantify 
their spectral shapes.

\section{Discussion}\label{discussion}

We have presented first observations of the cluster A4038 with the 
uGMRT in bands 3 (300 - 500 MHz) and 4 (1050 - 1450 MHz). The cluster has a 
central radio source associated with the BCG and a peculiar, extended ultra- 
steep spectrum radio relic located adjacent to the core. The 
rich spatial and spectral structure in the radio relic is analyzed using the 
uGMRT observations.

\subsection{Equipartition magnetic field}
Direct estimate of magnetic fields towards such relics is difficult. Using the 
measurements in radio bands, the magnetic field assuming equipartition can be 
calculated. Following, \citet{gov04}, the minimum energy density $u_{\rm min}$ is 
given by,
\begin{eqnarray}
 u_{\rm min}\Bigl[\frac{\mathrm{erg}}{\mathrm{cm}^3}\Bigr] = \xi(\alpha,\nu_1,\nu_2) 
(1+k)^{4/7} 
 (\nu_{0[\mathrm{MHz}]})^{4\alpha/7} \times \nonumber \\ (1+z)^{(12+4\alpha)/7}
  \bigl(I_{0[\frac{\mathrm{mJy}}{\mathrm{arcsec}^2}]}\bigr)^{4/7} 
(d_{[\mathrm{kpc}]})^{-4/7}
\end{eqnarray}
where $k$ is the ratio of energy in relativistic protons to that in electrons, 
$\alpha$ is the synchrotron spectral 
index, $\nu_0$ is the 
frequency at which the surface brightness, $I_{0}$ is measured, $d$ is the 
depth of the source and $\xi(\alpha,\nu_1,\nu_2)$ is a parameter that is a 
function of the spectral index 
and the lower and higher limits in frequency, $\nu_1$ and $\nu_2$ 
\citep{gov04}. In this formulation, the $K-$correction is included and a 
filling factor of 1 is assumed. The magnetic field is then 
given by, 
\begin{equation}\label{beq1}
 B_{\rm eq[\mathrm{G}]} = \Bigl(\frac{24 \pi}{7}u_{\rm min}\Bigr)^{1/2}.
\end{equation}
For A4038 relic we have $d= 80$ kpc which is the average of the projected 
extents 102 and 56 kpc.  At $\nu_0 = 402.0$ MHz we measure $I_{0} = 
0.0829$ mJy arcsec$^{-2}$ for A4038 relic. 
The modified equipartition magnetic field ($B^\prime_{\rm eq}$) based 
on a limit on the minimum Lorentz factor, $\gamma =100$ rather than on the 
frequency 
\citep[e.g.][]{1997A&A...325..898B, bec05} is given by, 
\begin{equation}
 B^{\prime}_{\rm eq[\mathrm{G}]} \sim 1.1 
\gamma^{\frac{1-2\alpha}{3+\alpha}}_{\mathrm{min}} 
B_{\rm eq}^{\frac{7}{2(3+\alpha)}},
\end{equation}
where $B_{eq[\mathrm{G}]}$ is from Eq.~\ref{beq1}. 

\begin{figure}
 \begin{center}
\includegraphics[height=7cm]{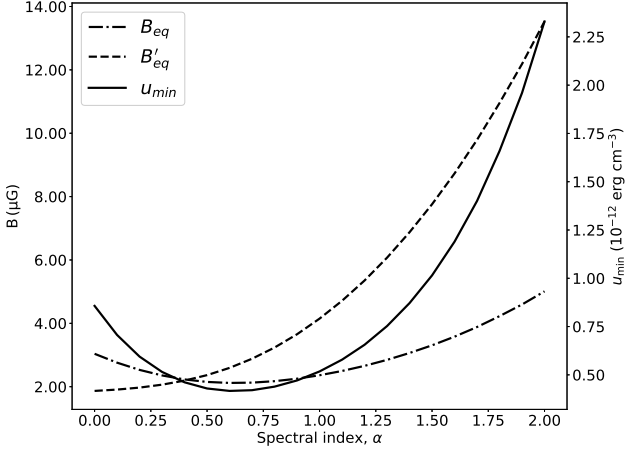}
\caption{The minimum energy and magnetic fields versus the integrated spectral index are shown. For the relic A4038 the integrated spectral index is 1.5 at the low frequency end. The magnetic field under the assumption of equipartition using fixed frequency interval ($B_{\rm eq}$) and fixed minimum Lorentz factor ($B^\prime_{\rm eq}$) are 3.3 and 7.8 $\mu$G, respectively.
\label{bequi}}
\end{center}
\end{figure}

The minimum energy and the magnetic fields determined for A4038 parameters as a function of the spectral index are shown in Fig.~\ref{bequi}. 
For A4038 the integrated spectral index in the range 0.074 - 0.327 GHz is 1.5 and steepens at higher frequencies \citep{kaldwa12}. 
At $\alpha=1.5$ we find the $\rm B_{\rm eq}$ and $\rm B^{\prime}_{\rm eq}$ to be 3.3 and 7.8 $\mu $G, respectively.

In the above formalism it has been assumed that the spectrum is a single power-law across the range of frequencies. However the integrated spectrum of A4038 relic is not a single power-law. Moreover the spectra are curved also within the relic. Therefore the standard formalism is insufficient to calculate the magnetic field in such systems. The occurrence of curvature in the spectrum may be due to superposition of distinct power-law spectra within the smallest region considered here to measure the spectrum. This can be checked with  more sensitive and higher resolution observations. A radically different  possibility is of an intrinsically curved electron energy distribution  that has been proposed to explain the curved spectra of radio galaxy lobes \citep{2012MNRAS.421..108D}.

\subsection{Relic and X-ray connection}
\begin{figure}
 \begin{center}
\includegraphics[height=9cm]{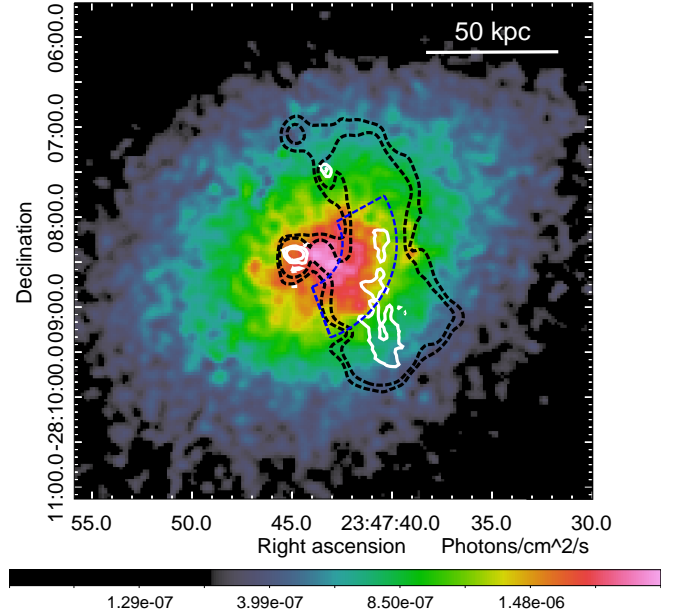}
\caption{The X-ray image smoothed to $5''$ resolution is shown in color 
scale. The dashed black contours are at 402 MHz (1.0 and 2.0 mJy beam$^{-1}$). 
The white contours are at 1.26 GHz (0.18 and 1.0 mJy beam$^{-1}$). The blue 
dashed region marks the arc-shaped extension of the {\bf central} bright elongated 
region in the X-ray image. The Arc region of the relic appears to skirt the 
X-ray arc and the Bridge is cospatial with the central elongated region in  
X-rays.
\label{xray}}
\end{center}
\end{figure}

In the case of diffuse radio sources it is non-trivial to make the association 
with either the X-ray emission or any optical counterpart. For A4038 relic 
we examined the evidence for connection with the ICM of the cluster. The 
BCG in the cluster (IC5358)  is a radio galaxy located at an offset of 
10 kpc from the peak in X-rays. The brightest region of X-rays is elongated in 
the northeast - southwest direction for 20 kpc, and the Bridge is 
co-spatial with it. The Arc region of the relic skirts a curved region in 
X-rays that extends to the north as shown in Fig.~\ref{xray}. The steeper 
northern parts of the relic 
extend beyond the X-ray arc. From the morphology, we infer that the 
Arc of the relic is near the cluster core and the rest of the relic  
extends beyond it. As noted by \citet{sle01}, there appears to be no 
connection between the optically detected galaxies and the relic. 

A spectral index map between 606 and 240 MHz with 
a resolution of $25''$ was presented in \citep{kaldwa12}. However that did not 
resolve the Arc and the Loop.
Here the spectral curvature across the relic is resolved using the uGMRT 
observations. We find that the integrated spectrum of the relic is not made up 
of self-similar spectral regions across the relic but shows significant 
variation in curvature (Fig.~\ref{spec}, right). The high frequency spectral 
index shows low variation though overall it 
is steep; it is the low frequency spectral index that shows flattening in the 
Arc region. Curved spectra and filamentary morphology are strong signatures for 
an origin in compression \citep{2002MNRAS.331.1011E}. The high curvature at the 
Arc region implies proximity of it to the disturbance that caused the 
compression. The presence of the Bridge (region 14) region supports a scenario 
where there is a connection between the relic and the BCG. The relic may have 
originated from the cocoon of the past activity of the BCG. 

There is evidence for the disturbance at the cluster centre from the 
detection of turbulent broadening of spectral lines in X-rays. Turbulent 
broadening in line widths of 1300 km s$^{-1}$ at the cluster core has been  
reported \citep{2011MNRAS.410.1797S}, which implies that it does not have a 
cool core. This supports the possibility of the disturbance at the core that 
displaced the BCG away from the X-ray peak and led to the compression of the 
radio plasma. A deeper investigation of the X-ray emission is required to find 
morphological features that correlate with the radio relic.

\subsection{Comparison with other relics}
Remnants of radio galaxies in general are not limited to be in galaxy clusters. 
From the samples of radio sources with steep spectra, possible 
remnants of double radio sources have been discussed
\citep{dwa09,wee11b,2015A&A...583A..89S}. Such remnants need not necessarily be 
steep spectrum sources \citep{2017A&A...606A..98B}. Several clusters with 
shock-related radio relics and central radio halos show presence of other 
smaller scale diffuse emission \citep[e. g.][]{2016ApJ...818..204V}. In order 
to carry out a fair comparison between relics, a sample of objects with 
possible common origin are needed. Here we specifically discuss remnant sources 
that have been found in confirmed clusters of galaxies that contain ICM 
detected in X-rays and do not have other large-scale complex diffuse radio 
sources. We selected our sample of cluster remnant relics starting from the 
samples of relics presented in \citet{fer12},\citet{wee11,wee11b} and 
\citet{nuza17}. Our sample is presented in Table~\ref{litrelics}. 
The integrated spectral indices in low and high frequency bands are compiled 
from the literature. 
The relics in A1367 and A610 are not considered due to lack of spectral and 
morphological information. A2256 relic may have contributions due to  
 direct shock acceleration at the relic and is thus excluded \citep{tra15}. 
According to the classification by \citet{nuza17}, there are five confirmed 
phoenixes, namely, A85, A2048, A2443, A4038 and 24P73. We excluded the relic 24P73 due to the absence of confirmed diffuse X-ray emission from the 
associated galaxy cluster. 

The remnant relics are typically irregular in appearance. The mean distance of the sample remnant relics from their host cluster centres is 0.36 Mpc, with A4038 being the one closest to the cluster centre. 
The morphology and spectra 
of these relics indicate a high chance that these are revived remnants or 
phoenixes. The host clusters are all nearby, ranging in redshifts from 0.01- 
0.13 and have a mean redshift of 0.07. This is not surprising as the lifetimes of the remnants would reduce with 
increasing redshift due to the inverse Compton losses to Cosmic 
Microwave Background photons ($t_{\rm loss} \propto (1+z)^{-4}$). Since the sample 
is not from a complete sample of clusters across redshifts, the occurrence of 
such relics cannot be judged yet. The location of 
the relics relative to the cluster 
centers is in the range 0.02 - 0.5 Mpc and shows no trend with redshift; A1664 
is the only outlier at a distance of $1$ Mpc from the cluster centre. 

\begin{figure}
 \begin{center}
\includegraphics[height=8cm]{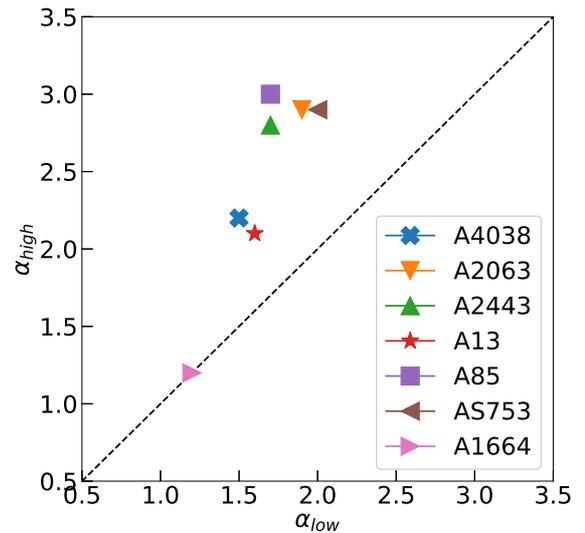}
\caption{The high versus low frequency spectral indices for the remnant relics sample described in Table~\ref{litrelics} are shown. 
\label{color}}
\end{center}
\end{figure}

From the integrated spectra reported in the literature, the low and high 
frequency spectral indices were determined and are reported in 
Table~\ref{litrelics}. We do not find any trend between the distance of the 
relic from the cluster centre and the spectral indices or the curvatures. 
The curvatures are all greater 
than 0.5 with A85 relic showing the highest curvature of 1.3. 
In Fig.~\ref{color}, the low and high frequency spectral indices 
 for the sample relics derived from the integrated spectra are shown. The deviation above the 
dashed line ($\alpha_{\rm low} = \alpha_{\rm high}$) indicates a positive curvature 
in the spectrum. The relic A1664 is an outlier among the considered relics with the flattest spectra. The rest of the sample has spectra steeper than 1.5 and are thus either in the passively evolving remnant phase or have been revived by mechanisms other than direct injection by an AGN. These mechanisms include adiabatic compression by ICM shocks and gentle re-acceleration processes \citep{ens01,2017NatAs...1E...5V}. The spectral indices across the A4038 relic 
presented with wideband observations show values steeper than 1.5 and a variation in the range 1.2 - 3.7. The integrated spectrum is thus composed of low and high brightness regions with a wide variety of spectral curvatures that are not well represented by the curvature of the integrated spectrum.
It is essential to carry out 
resolved studies such as the one presented for A4038 relic here. The uGMRT is an ideal instrument to carry out these studies for samples of relics. 
Studies with closely matched observations between X-rays and radio are needed in order to find the phenomena behind the formation of the remnant relics in galaxy 
clusters.

\section{Summary and Conclusions}\label{conclusions}
We have presented a spectral curvature study of the relic in Abell 4038 using 
the 
uGMRT in the 300 - 500 MHz and 1050 - 1450 MHz bands. The connection between 
the 
relic properties and the X-ray emission from the cluster was discussed. A  
comparison between the A4038 relic and other small scale cluster relics was 
also carried out. The main findings are summarized below:
\begin{enumerate}
 \item We have produced deep images of the radio relic in A4038 at 0.402 and 
1.26 GHz using the uGMRT with bandwidths of 200 and 400 MHz in the two bands, 
respectively. The rms noise at the centre of the field at 402 MHz was $0.07$ 
mJy beam$^{-1}$ and at 1.26 GHz was $0.03$ mJy beam$^{-1}$.
 The largest extent of the relic in the north-south direction is 102 kpc 
and in the east-west direction is 58 kpc. 
\item The spectra across the relic were estimated in 15 regions, each 
of size $15''\times15''$.
The spectrum of each region was fit with separate power-laws in the ranges 0.32 
- 0.45 and 0.45 - 1.43 GHz. The difference in the spectral indices in the low 
($\alpha_{\rm low}$) and high frequency ($\alpha_{\rm high}$) ranges was used as a 
measure of the spectral curvature ($\Delta \alpha$).
\item  The highest curvature, $\Delta \alpha$, of $1.6\pm0.3$ was found in the 
region corresponding to the Arc. The curvature is 0.8 in the regions 10 and 11 that make the eastern and southern parts of the Loop, respectively. To the west of the Loop it becomes consistent with zero in region 12 and then rises to 0.5 further west.
\item A curved arc-like region in the X-rays is skirted by the Arc 
in the relic. We propose that the highly curved spectra in the Arc result from 
compression-revived radio emission. This is consistent with our earlier work 
presented in \citet{kaldwa12}, where the integrated spectrum was best fit in 
the adiabatically compressed phase in the model by \citet{ens01}.
\item The calculation of magnetic field under the minimum energy condition is dependent on the spectral index and the current formalisms assume a single power-law. For A4038 relic if we assume a spectral index of 1.5, the magnetic field is 7.8 $\mu$G if the minimum Lorentz factor, $\gamma_{\rm min}=100$.
\item We also present a sample of 10 remnant relics in galaxy clusters that 
have detectable X-ray emission from their ICM. These relics are typically found 
within the 0.5 Mpc from the cluster centre-the only exception being A1664. 
\item We have plotted low versus high frequency spectral indices for the remnant relic sample, together with the values for the A4038 relic. Except the A1664 relic which has a flat and straight spectrum, the sample remnant relics have curvatures in the range 0.5 - 1.6. These are consistent either with an old synchrotron plasma or a plasma re-energised by a mechanism such as adiabatic compression by shock or gentle re-acceleration.
\item We conclude that for A4038 the spectral shape changes across the regions within the relic and thus spatially resolved spectral studies as presented for the case of A4038 relic are essential to study the origins of such relics.
\item The deep radio images from the uGMRT presented here also resulted in 
the detection of faint lobes around a radio core identified with the galaxy 
 PGC072471.  
\end{enumerate}

\section*{Acknowledgements}
We thank the anonymous referee for critical comments that have improved this paper. RK acknowledges support through the DST-INSPIRE 
Faculty Award by the Government of India.
We thank the staff of the GMRT that made these observations possible. 
GMRT is run by the National Centre for Radio Astrophysics of the Tata Institute 
of Fundamental Research. This research has made use of 
the NASA/IPAC Extragalactic Database (NED) which is operated by the Jet 
Propulsion Laboratory, California Institute of Technology, under contract with 
the National Aeronautics and Space Administration. This research made use of 
Astropy, a community-developed core Python package for Astronomy (Astropy 
Collaboration, 2018). The scientific results reported in this article are based 
in part on data obtained from the Chandra Data Archive.

\begin{table*}
\begin{center}
\caption{\label{fit} Spectral properties of the regions (Sec~\ref{curv}).}
\begin{tabular}{ccccccc}
\hline\noalign{\smallskip}
Region & $\alpha_{\rm low}$ & B & $\alpha_{\rm high}$ & C & $\Delta \alpha$ \\
number &  &  &  &  & ($\alpha_{\rm high}-\alpha_{\rm low}$) \\
\hline\noalign{\smallskip}
0 & $ 2.6 \pm 0.5 $ & $ 20.2 \pm 4.0 $ &- &-&-\\
1 & $ 3.7 \pm 0.5 $ & $ 29.2 \pm 4.7 $ & - & -&-\\
2 & $ 2.6 \pm 0.6 $ & $ 19.7 \pm 5.3 $ &- & -&-\\
3 & $ 2.1 \pm 0.6 $ & $ 16.0 \pm 4.8 $ & - & -&-\\
4 & $ 2.2 \pm 0.8 $ & $ 16.7 \pm 6.9 $ & $ 3.1 \pm 0.1 $ & $ 24.4 \pm 1.1 $&$  
0.9 \pm 0.8 $\\
5 & $ 1.9 \pm 0.1 $ & $ 14.4 \pm 1.2 $ & $ 2.8 \pm 0.2 $ & $ 22.7 \pm 1.5 $&$  
0.9 \pm 0.2 $\\
6 & $ 1.9 \pm 0.1 $ & $ 15.0 \pm 1.0 $ & $ 2.8 \pm 0.2 $ & $ 23.0 \pm 1.5 $&$  
0.9 \pm 0.2 $\\
7 & $ 1.4 \pm 0.1 $ & $ 10.5 \pm 1.0 $ & $ 2.7 \pm 0.2 $ & $ 22.3 \pm 1.6 $&$  
1.4 \pm 0.2 $\\
8 & $ 1.2 \pm 0.2 $ & $ 9.2 \pm 2.1 $ & $ 2.8 \pm 0.2 $ & $ 23.0 \pm 2.2 $&$  
1.6 \pm 0.3 $\\
9 & $ 1.6 \pm 0.2 $ & $ 12.1 \pm 1.7 $ & $ 3.0 \pm 0.2 $ & $ 24.5 \pm 2.0 $&$  
1.3 \pm 0.3 $\\
10 & $ 1.7 \pm 0.2 $ & $ 13.3 \pm 1.6 $ & $ 2.5 \pm 0.2 $ & $ 19.5 \pm 2.1 $&$ 
 0.8 \pm 0.3 $\\
11 & $ 1.8 \pm 0.2 $ & $ 14.1 \pm 2.0 $ & $ 2.6 \pm 0.2 $ & $ 21.7 \pm 2.1 $&$ 
 0.8 \pm 0.3 $\\
12 & $ 2.3 \pm 0.2 $ & $ 18.5 \pm 1.7 $ & $ 2.3 \pm 0.2 $ & $ 18.4 \pm 2.2 $&$ 
 -0.0 \pm 0.3 $\\
13 & $ 2.1 \pm 0.2 $ & $ 16.5 \pm 1.7 $ & $ 2.6 \pm 0.2 $ & $ 20.5 \pm 2.1 $&$ 
 0.5 \pm 0.3 $\\
14 & $ 2.2 \pm 0.5 $ & $ 16.5 \pm 3.9 $ & $ 2.6 \pm 0.3 $ & $ 20.4 \pm 2.3 $&$ 
 0.4 \pm 0.5 $\\
\hline\noalign{\smallskip}
\end{tabular}
\end{center}
\end{table*}

\begin{table*}
\centering

\caption{\label{litrelics} Summary of remnant relics from the literature. Col. 
1: 
Cluster name; Col. 2: Redshift; Col. 3: Distance from the cluster center; Col. 
4: Low frequency spectral index with the frequency range given in GHz; Col. 5: 
High frequency spectral index with the frequency range given in GHz; Col. 6: 
Curvature  defined as the difference between the high and low frequency 
spectral indices. Note: The typical error on the reported spectral indices is 
 0.1 - 0.2.}
 \begin{tabular}{ccccccccc}
 
 \hline\noalign{\smallskip}
 
Name	 &z & Rcc & $\alpha_{\rm low}$ &$\alpha_{\rm high}$& $\Delta \alpha$& 
Reference\\

	 & & (Mpc) & (GHz, GHz) &(GHz, GHz)& Curvature&\\
	 
\hline\noalign{\smallskip}

{\bf A4038}&0.02819	 &0.02	&1.5 (0.074, 0.327) & 2.2 (0.327, 
1.4)&0.7& \citet{kaldwa12}\\

A2063	 &0.0349	&0.04	&1.9 (0.08, 0.408) &2.9 (0.408, 
1.465)&1.0& \citet{KG94}\\

A2443	 &0.1080	&0.23	&1.7 (0.074,0.325) & 2.8 (0.325, 
1.4)&1.1&\citet{cohen11}\\
A13	 &0.0943	&0.2	&1.6 (0.08, 0.327)  & 2.1 
(0.327,1.4)&0.5&\citet{sle01}\\
A85	 &0.0551	&0.43	&1.7 (0.08,0.327)   & 3.0 (0.327, 1.4)& 
1.3&\citet{sle01}\\
A548b-NW &0.0424	&0.43	&-&$\sim$2 (1.36, 
14.3)&-&\citet{2006MNRAS.368..544F}\\
A548b-N	 &0.0424	&0.5	&-&$\sim2$ (1.36, 
14.3)&-&\citet{2006MNRAS.368..544F}\\
AS753	 &0.014		&0.41	&2.0 
(0.33,1.398)&2.9 (1.398,2.378)&0.9&\citet{2003AJ....125.1095S}\\
A1664	 &0.1283	&1.03	&1.2 (0.15,1.4)&1.2 
(0.15,1.4)&$0.0$&\citet{kaldwa12}\\
A2048	 &0.0972	&0.33	&-(0.074,0.610)&-(0.61,1.4)&1.6&\citet{wee11}\\
\hline\noalign{\smallskip}
\end{tabular}
\end{table*}

\begin{figure*}
 \begin{center}
 \includegraphics[height=6cm]{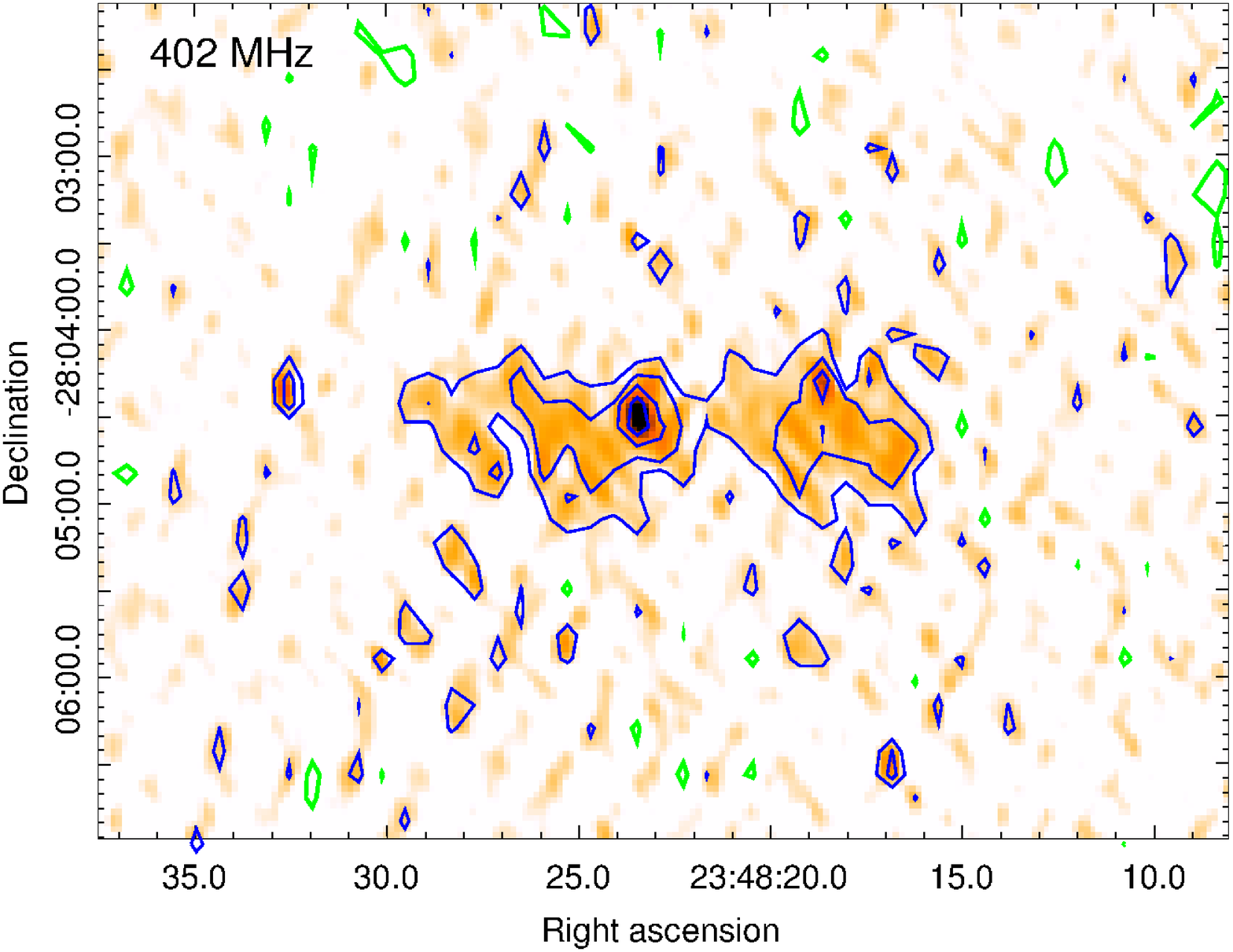}
  \includegraphics[height=6cm]{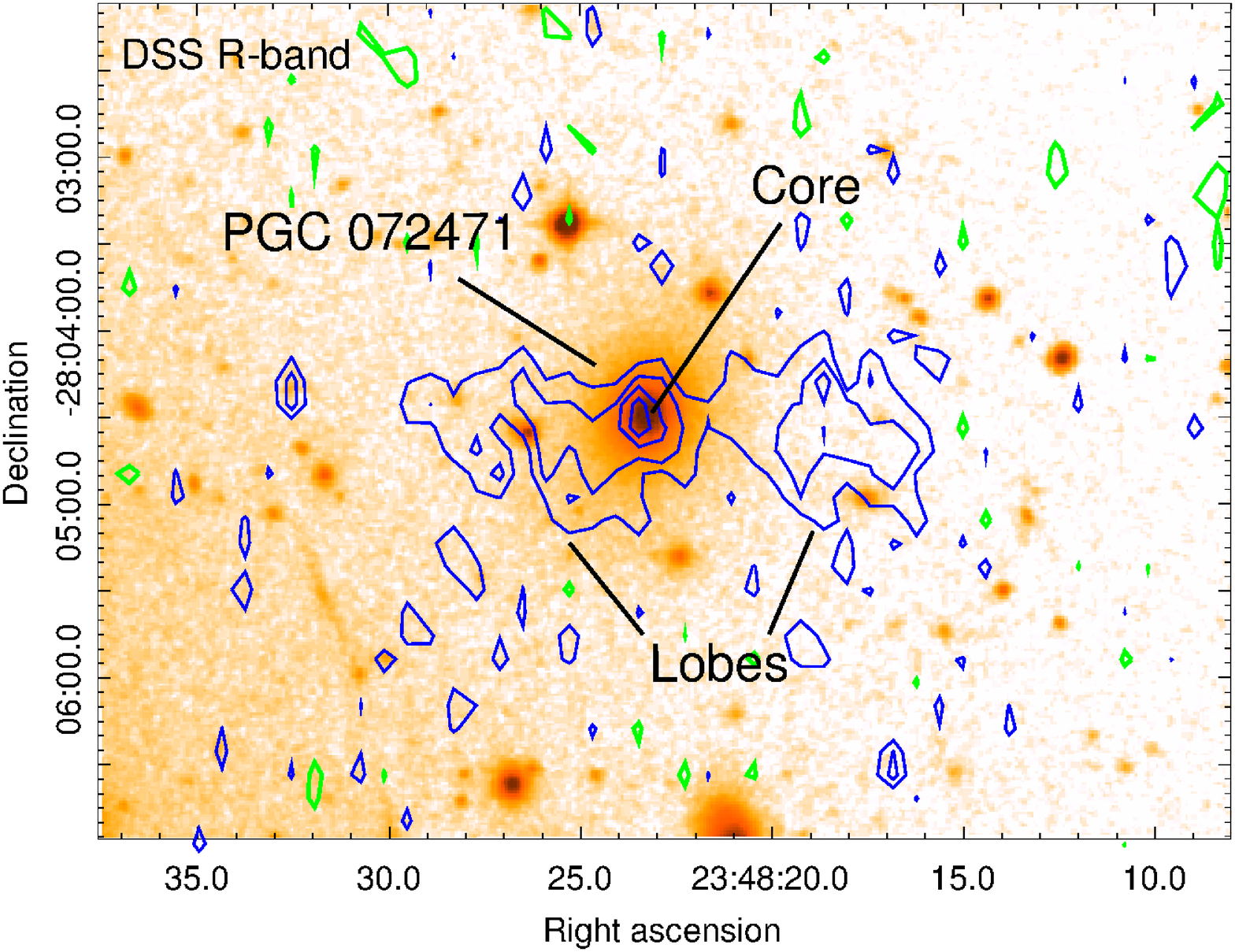}
\caption{Left:- The 402 MHz image zoomed-in on the rectangle (unrelated to the A4038 relic) in Fig.\ref{imgpl} is shown in color and contours. The 
contours are at -0.1, 0.1, 0.2, 0.4,... mJy beam$^{-1}$. The positive 
contours are blue and the negative are green. Right:- The Digitized Sky Survey 
R-band image is shown in color with the contours from the left panel overlaid. 
The compact core can be identified with the galaxy PGC 072471 detected in 
optical. The lobes are detected for the first time.
\label{fig:imgapp}}
\end{center}
\end{figure*}



\bibliographystyle{mnras}
\bibliography{mybib_1} 





\appendix

\section{Detection of diffuse lobes of a radio galaxy}\label{app}
We have detected faint diffuse lobes around a compact source in our uGMRT 402 
MHz image. The compact source is associated with the galaxy PGC072471 that is 
located at the redshift of $0.034130\pm0.000067$ \citep{1996A&AS..116..203S}. 
The diffuse lobes extend nearly symmetrically on the east and west of the 
compact source. The optical and radio images are shown in Fig.~\ref{fig:imgapp}.
The compact core has a flux density of 2.6 mJy and total flux density 
including the lobes is 26.6 mJy. These are not corrected for the effect of the 
primary beam. The distance of this source from the phase centre is $10'$. The 
east-west extent of the lobes is 193'' which is equal to 109 kpc at the 
redshift of the host galaxy.


\bsp	
\label{lastpage}
\end{document}